\documentclass[aps,pra,amsmath,twocolumn, nofootinbib,showpacs,superscriptaddress]{revtex4}
\usepackage{epsfig}
\usepackage{color}
\usepackage{psfrag}

\newcommand{\ket}[1]{\vert #1 \rangle}
\newcommand{\bra}[1]{\langle #1 \vert}
\newcommand{\bracket}[2]{\langle #1 \vert #2 \rangle}

\newcommand{\tr}{\mathop{\mathrm{tr}}\nolimits}

\newcommand{\RePart}{\textrm{Re}}
\newcommand{\ImPart}{\textrm{Im}}

\newcommand{\psz}{|\psi\rangle}
\newcommand{\pt}{|\psi_\theta\rangle}
\begin{document}
\title{Phase variance of squeezed vacuum states}
\author{E. Bagan}
\affiliation{Departament de F\' isica, Universitat Aut\`onoma de
Barcelona, Bellaterra E-08193, Spain}
\author{A. Monras}
\affiliation{The School of Physics and Astronomy, University of
Leeds, Leeds LS2 9JT, United Kingdom}
\author{R. Munoz-Tapia}
\affiliation{Departament de F\' isica, Universitat Aut\`onoma de
Barcelona, Bellaterra E-08193, Spain}

\begin{abstract}
 We consider the problem of estimating the phase of squeezed vacuum
states within a Bayesian framework. We derive bounds on the average
Holevo variance for an arbitrary number $N$ of uncorrelated copies.
We find that it scales with the mean photon number, $n$, as dictated by the
Heisenberg limit, i.e., as~$n^{-2}$, only for $N>4$. For $N\leq 4$ this
fundamental scaling breaks down and it becomes  $n^{-N/2}$.
Thus, a single squeezed vacuum state performs
worse than a single coherent state with the same energy. We find
the optimal splitting of a fixed given energy among various copies.
We also compute the variance for repeated individual measurements
(without classical communication or adaptivity) and find that the
standard Heisenberg-limited scaling $n^{-2}$ is recovered for large samples.
\end{abstract}
 \maketitle

\section{Introduction}

Squeezed states can improve the sensitivity of laser interferometry
with a precision such as to beat the so-called shot-noise
limit~\cite{bachor} and, therefore, have been considered as useful
states in a wide variety of applications. They offer an enhanced
resolution-energy tradeoff as compared to coherent state
interferometry~\cite{xaowokimble87,kobolov,caves}. Some recent
applications include precision measurements of
distances~\cite{treps2}, detection of small displacements in optical
images~\cite{fabre}, or optical imaging~\cite{treps} with multimode
light. One of the most promising fields of application is the
detection of gravitational waves. This idea has been recurrently
discussed in the last two decades~\cite{caves} and it is finally
being included in the latest experimental
proposals~\cite{kimble,ligo,vahlbruch}. Extensions to non-optical
systems such as squeezed atomic states have also been
considered~\cite{lukin, monroe, molmer}.

The main advantage of squeezed states is that they can have optical
phase variance below the standard quantum limit. In this context the
most relevant ones are the squeezed vacuum states
(SVS)~\cite{amonras06}. It is generally claimed that these states
have a phase variance scaling as $n^{-2}$, where $n$ is their
average photon number. This result is hinted at by a somewhat
heuristic argument that uses the Heisenberg uncertainty relation,
and so, this $n^{-2}$ scaling is known as the Heisenberg Limit (HL).
A rigorous proof was derived by Holevo~\cite{holevo84, holevo84b},
and can also be obtained from the Cramer-Rao bound
\cite{cramer46} and the Braunstein-Caves information inequality
\cite{braunsteincaves94}. This work served as the foundation for
recent developments in entanglement-enhanced metrology~\cite{giovanetti, metrology}.
The bounds on the variance derived
in~\cite{amonras06,holevo84, holevo84b,cramer46,braunsteincaves94,giovanetti, metrology},
however, are generally tight
only when a large number of independent trials are repeated, whereas
for small number of trials the attainability is not guaranteed.


Our aim here is to determine the attainable precision in phase
estimation for an arbitrary number $N$, \emph{not necessarily large}, of
uncorrelated, identically prepared SVS. To the best of our knowledge, no
study has yet addressed this issue.
Our analysis may be relevant in situations of quantum-limited communication
in which one has access to few states
(see~\cite{schleich93}) or in  gravitational wave detection, where
the available exposure time is limited and a large sample cannot be
assumed.
To tackle this situation we adopt a Bayesian approach, in
which an averaged cost function is minimized over all possible
estimation strategies. For the problem at hand, the Holevo phase
variance~\cite{holevo84} is particularly well suited. This approach
will enable us to perform a complete analytical computation and
derive closed expressions for the optimal averaged phase variance.
We pay particular attention to the asymptotic values of this
variance for SVS with large and small mean photon number. We also
study the large $N$ regime and compare the values with those for
repeated individual measurements.

We obtain the rather surprising result that the  scaling~$n^{-2}$ of
the Heisenberg limited variance of~$N$ SVS can only be attained if $N$ is greater than four, while for $1\leq N\leq 4$ the scaling is given by
$n^{-N/2}$ (see Fig.~\ref{fig:loglog} below). This shows that
a single SVS cannot do better than a single coherent state with
the same energy. For large~$N$, we, of course, recover the HL expression. Our analysis also shows that for a given amount of energy $E$ that can be split among $N$ identical SVS states of mean photon number $n$, i.e. $E=nN$ (in units of $\hbar\omega$), the minimum variance is not attained with a single state, $N=1$, of maximal mean photon number, as the $n^{-2}$ scaling would suggest. For a large amount of available energy,  we find that the optimal choice consists of splitting the energy among $N=8$ identical copies (see Fig.~\ref{fig:N8} in Sec.~\ref{sec:squeezed}).

In this work we consider the most general measurements for
optimizing the resolution and, therefore, our results can be
regarded as the maximum precision limits to phase estimation with
SVS allowed by quantum mechanics. In this sense, our results are also
relevant for metrology.  Although there exist some quantum metrology precision
bounds~\cite{metrology} that can  surpass the HL, they require the use of nonlinear
Hamiltonians, which are in general very difficult
to implement.

The paper is organized as follows. In the next section we present
the basics of our approach by considering the simpler case of
coherent states. This serves as a warm-up exercise and as an
illustration of the main techniques used throughout this paper. In
Section~\ref{sec:squeezed} we move to the core of our work. We
derive the optimal measurement for an arbitrary number of identical
copies of SVS and~we obtain the bound on the phase variance. In
Section~\ref{sec:individual} we provide results for non-adaptive
individual measurements. In Section V we discuss the implications of
our results and draw our conclusions. The paper ends with two
technical appendices.


%
%
%
%
%

\section{Coherent states} \label{sec:coherent}
In this section we introduce the basic elements of our approach and
consider the simpler case of coherent states as an example. These
states are given by $\psz=D(\alpha)\ket{0}$, where $D(\alpha)$ is
the displacement operator $D(\alpha)=\exp[\alpha
a^\dagger-\alpha^{*} a]$ and $a$ ($a^\dagger$) are the standard
photon anihilation (creation) operators of a generic mode. The state
of $N$ identical copies of $|\psi\rangle$ is simply
$\ket{\Phi}=\psz^{\otimes N}$ or, equivalently,
$\rho=\ket{\Phi}\bra{\Phi}$. The phase shifted state $\pt$ is the
result of acting on $|\psi \rangle$ with the unitary operator
$U(\theta)=\exp[i\theta a^\dagger a]$ (e.g., the time evolution
operator for the free electromagnetic mode: $\theta=\hbar\omega t$),
i.e.,  $\pt=U(\theta)\psz$. The set of $N$ identical shifted states
are likewise written as  $\ket{\Phi_{\theta}}=\pt^{\otimes N}$ or
$\rho_{\theta}=\ket{\Phi_{\theta}}\bra{\Phi_{\theta}}$.

Since our purpose is to estimate $\theta$ with highest precision, we
allow ourselves to perform the most general measurements on
$|\Phi_\theta\rangle$. These, so-called generalized measurements,
are described by a Positive Operator Valued Measure (POVM); that is,
a set $\mathcal M=\{O_\chi\}$ of positive operators, $O_\chi\geq0$,
that add up to the identity, $\sum_\chi O_\chi=\openone$. Also, for
each outcome of the measurement we need to give a guess of the value
of $\theta$, in technical words, we have to choose an appropriate
estimator~$\hat\theta:\chi\mapsto\hat\theta_\chi$. In the Bayesian
approach this is done by providing a  cost function (figure of
merit) for the estimation protocol. The optimal choice of both
measurement and estimator is taken to be that that minimizes
(maximizes) the averaged cost function (figure of merit). We take
the cost function
\begin{equation}
    \label{eq:HolPV}
    V=|\langle e^{i(\theta-\hat\theta)}\rangle|^{-2}-1,
\end{equation}
which is the natural variance for cyclic variables, as pointed out
by Holevo~\cite{holevo84}. Note that $V$ vanishes for perfect
estimation, and goes to infinity for a flat distribution of
$\hat\theta$ (random guessing). Notice also that the minimization of
the variance $V$ does not guarantee that $\hat\theta$ is close to
$\theta$ unless some sort of unbiasedness condition is imposed. For
the distributions in the circle considered here the following
conditions are sufficient:
\begin{equation}
    \label{eq:unbiased}
    \ImPart\langle e^{i(\theta-\hat\theta)}\rangle=0 , \quad
    \RePart\langle e^{i(\theta-\hat\theta)}\rangle\geq 0.
\end{equation}
We find it also convenient to introduce the figure of merit
$F=|\langle e^{i(\theta-\hat\theta)}\rangle|$, which we will loosely
refer to as {\em fidelity}. This name is suggested by the fact that
$0\leq F\leq 1$, with the  values 0 and 1 for the completely random
guessing and perfect estimation, respectively. The relation with the
Holevo variance can be read off from Eq.~\eqref{eq:HolPV}:
$F=(1+V)^{-1/2}$.  The average in $F$ (or $V$) is over all possible
outcomes as well as all possible signal states,
\begin{equation}
    F=\left|\sum_\chi\int\frac{d\theta}{2\pi}
   e^{i(\theta-\hat\theta_\chi)}\tr[O_\chi
    \rho_\theta]\right|
    \label{eq:generalFid},
\end{equation}
where we have assumed a flat prior distribution on the circle. This
is the natural choice when nothing is known beforehand about the
phase we wish to estimate. An additional feature of the Holevo phase
variance $V$ is that it approaches the statistical variance
$(\Delta\hat\theta)^2\equiv\langle\hat\theta^2\rangle-\langle\hat\theta\rangle^2$
in the limit of accurate estimation (peaked distributions), i.e,
\begin{equation}
    \label{eq:VFDcoh}
    V\simeq2(1-F)\simeq (\Delta\hat\theta)^2.
\end{equation}

To maximize $F$, it is useful to write the coherent state in the
photon number eigenbasis (also referred to as Fock basis):
$\rho_{\theta}=\sum_{kl}e^{i\theta(k-l)}\rho_{kl}\ket{k}\bra{l}$. It
is not difficult to prove that $F$ is bounded by (see~\cite{bagan05}
for details)
\begin{multline}
     F=\left|\sum_\chi e^{-i\hat\theta_\chi}\sum_{k}\rho_{k\,k+1}[O_\chi]_{k+1\, k}\right|\\
     \leq\sum_k|\rho_{k\,k+1}|\sum_\chi\left|[O_\chi]_{k+1\, k}\right|,
\end{multline}
which becomes an equality if $[O_\chi]_{k+1\, k}=|[O_\chi]_{k+1\,
k}|e^{i\hat\theta_\chi}$ for all $k$ and $\chi$. Applying the
Cauchy-Schwartz inequality and using positivity and the completeness
of $O_\chi$, which together amount to
$\sum_\chi|[O_\chi]_{m,n}|\leq1$, we finally obtain
\begin{equation}\label{F-coherent}
    F\leq \sum_k|\rho_{k\,k+1}|.
\end{equation}

One can easily convince oneself that Holevo's canonical phase
measurement~\cite{holevo84}
\begin{equation}
\label{eq:generalMeas}
    [O(\phi)]_{k l}=\frac{1}{2\pi}e^{-i \phi(k-l)},
\end{equation}
saturates Eq.~\eqref{F-coherent} and satisfies the unbiasedness
condition~\eqref{eq:unbiased}~\cite{bagan05}.
Here, the  matrix elements are again written in the Fock basis
$\{\ket{k}\}$, $\phi$ is a uniform continuous parameter $\phi\in[0,2
\pi)$ that labels the outcomes, i.e., plays the role of $\chi$, and
the optimal estimator is simply given by $\hat\theta_\phi=\phi$,
with $\phi\in[0,2\pi)$.

At this point, the only task left is to compute the sum of matrix
elements in Eq.~\eqref{F-coherent}. Recall that a  coherent state
with mean photon number $n$  is given by
$\ket{\alpha}=\exp(-|\alpha|^2/2)\sum_k\alpha^k/\sqrt{k!}\ket k$, with
$|\alpha|=\sqrt{n}$. For a single copy of a coherent state it is
straightforward to obtain from Eq.~\eqref{F-coherent}   that the
maximum value of $F$ is
\begin{equation}
    \label{eq:stdCoh}
    F=e^{-|\alpha|^2}\sum_{k=0}^\infty\frac{|\alpha|^{2k+1}}{k!\sqrt{k+1}}.
\end{equation}
It proves useful to cast the above expression into an integral form,
which is  easier to study analytically. For this purpose, we use the
identity
\begin{equation}\label{integral-trick-1}
    \frac{1}{\sqrt{k+1}}=\frac{1}{\sqrt{\pi}}\int_0^\infty \frac{dt}{\sqrt
    t}\;e^{-t(k+1)}
\end{equation}
and perform the (now trivial) summation over $k$
in~\eqref{eq:stdCoh}. By an appropriate change of variables one gets
\begin{equation}
    \label{eq:IntegralCoh}
    F=\frac{|\alpha|}{\sqrt \pi}\int_0^1 dx \frac{e^{-x|\alpha|^2}}{\sqrt{-\log
    (1-x)}}.
\end{equation}
Eq.~\eqref{eq:IntegralCoh} is an integral representation of
Eq.~\eqref{eq:stdCoh} and can be easily computed to arbitrary precision for
any value of~$\alpha$.

The large~$n$ regime can now be worked out. Prior to
integration, we Taylor-expand the logarithm around~$x=0$ in
Eq.~\eqref{eq:IntegralCoh} and then integrate term by term. Note
that each power of $x$ that we retain gives a contribution of order
$1/|\alpha|^2=1/n$. The upper limit of the corresponding integrals
can be safely taken to be infinity, since this change will only
contribute to differences that fall off exponentially with~$n$. To
subleading order we obtain
\begin{equation}
    F=1-\frac{1}{8n}+\ldots\, .
\end{equation}
This is the maximum average fidelity attained with a single, highly
energetic coherent state ($n\gg1$). Using Eq.~\eqref{eq:VFDcoh} one
has
\begin{equation}\label{eq:stat-vari-coh}
    (\Delta\hat\theta)^2=\frac{1}{4n},
\end{equation}
which agrees with the well known statistical variance of coherent
states \cite{wiseman97}. The alternative derivation we have
presented here will prove very useful for~SVS, as will become
apparent in the remainder of the paper.

Interestingly, the same analysis can be carried out for an arbitrary
number of coherent states.
The case of two copies $N=2$ already contains all the ingredients of
the solution for arbitrary $N$. We recall that a symmetric state
with  total photon number $k$ is
\begin{equation}\label{eq:def-sym-coh}
    \ket{k}=\frac{1}{\sqrt{2^k}}\kern -0.5em \sum_{
     \begin{array}{c}
     \\[-1.6 em]
     \scriptscriptstyle n_1,n_2 \\[-0.5em]
     \scriptscriptstyle n_1+n_2=k
     \end{array}
    }\kern -0.5em \sqrt{\frac{k!}{n_1!\, n_2!}}\ket{n_1}\ket{n_2}.
\end{equation}
By using this definition we see that $|\alpha\rangle^{\otimes 2}$ is
unitarily  equivalent to a coherent state with amplitude $\sqrt
2\alpha$. More precisely,
\begin{eqnarray}
    \label{eq:TwoModeCohSt}
    \ket{\alpha}^{\otimes
    2}=e^{-|\sqrt2\alpha|^2/2}\sum_{k=0}^\infty\frac{(\sqrt2\alpha)^k}{\sqrt{k!}}\ket{k}.
\end{eqnarray}
Hence, two identical coherent states
$|\alpha\rangle_A|\alpha\rangle_B$ can be transformed by a two-mode
unitary transformation into a single coherent state
$\ket{\sqrt2\alpha}_C\ket0_D$ (this unitary can be simply realized,
e.g., by a 50/50 beam splitter). To simplify the notation, we drop
the mode labels ($A$, $B$,~\dots) and the the vacuum states
($\ket{0}$), as in~\eqref{eq:def-sym-coh} and
\eqref{eq:TwoModeCohSt}, throughout this paper. Applying this
reasoning inductively we see that an optimal generalized measurement
on $N$ identical coherent states $\ket\alpha^{\otimes N}$ is
formally equivalent to a single measurement on a
coherent state $\ket{\sqrt N\alpha}$. Thus, for large $N$ we get the
HL relation
\begin{equation}
    (\Delta\hat\theta)^2=\frac{1}{4nN}.
\end{equation}
\section{Squeezed Vacuum States}\label{sec:squeezed}
We next address the case of SVS. Although the calculations are
substantially more involved, the techniques are not so different
from those presented in the preceding section. Here $\pt$ is a SVS
given by $\pt=U(\theta)S(r)\ket 0$, where $S(r)$ is the squeezing
operator $S(r)=\exp[r(a^\dagger{}^2-a^2)/2]$, and $U(\theta)$ is the
phase shift operator already defined at the beginning of
Section~\ref{sec:coherent}. The parameter $r$ is usually referred to
as the squeezing parameter. The SVS in the Fock basis read
\begin{equation}\label{svs}
    \pt=(1-\beta^2)^{1/4}\sum_{k=0}^\infty\left(\frac{\beta e^{2i\theta}}{2}\right)^k\frac{\sqrt{
    (2k)!}}{k!}\ket{2k},
\end{equation}
where one can readily see that the SVS are superpositions of Fock
states, $\ket{2k}$, with an even number of photons. In
Eq.~\eqref{svs} we have defined $\beta=\tanh r$, which in turn is
related to the mean photon number through $\beta=\sqrt{n/(n+1)}$.

The $N$-copy state vector $\ket{\Phi_\theta}=\pt^{\otimes N}$ can be
written as
\begin{eqnarray}
    \label{eq:statePhi}
    \ket{\Phi_\theta}&=&(1-\beta^2)^{N/4}\sum_{k=0}^\infty (\beta e^{2i\theta})^kh_k\ket{2k},\\
    \label{eq:dks}
    h_k&=&\sqrt{\binom{N/2+k-1}{k}}\, ,
\end{eqnarray}
where $\ket{2k}$ are the $N$-mode symmetric states with total photon
number $2k$.
E.g., for $N=2$ they read
\begin{equation}\label{eq:sym-squeezed}
\ket{2k}=\frac{1}{2^k h_k} \sum_{
 \begin{array}{c}
     \\[-1.6 em]
     \scriptscriptstyle n_1,n_2 \\[-0.5em]
     \scriptscriptstyle n_1+n_2=k
     \end{array}
} \kern-0.5em \frac{\sqrt{(2n_1)!\, (2n_2)!}}{n_1!\,
n_2!}\ket{2n_1}\ket{2n_2},
\end{equation}
where in this case $h_k=1$. The generalization for arbitrary $N$ is
straightforward. Note that the explicit form of $h_k$ guarantees
that the normalization condition
$\bracket{\Phi_{\theta}}{\Phi_{\theta}}=1$ is fulfilled.
Actually,
\begin{equation}\label{eq:negative-binomial}
    \wp_k=h_k^2\, \beta^{2k} (1-\beta^2)^{N/2}
\end{equation}
can be viewed as the probability mass function of a negative binomial distribution~\cite{stirzaker}
with failure probability given by~$\beta^2$.

The invariance  $\ket{\Phi_{\theta+\pi}}=\ket{\Phi_{\theta}}$
 imposes a minor modification of the
Holevo phase variance~\eqref{eq:HolPV} for SVS, which now reads
\begin{equation}
    \label{eq:SVSvar}
        V=|\langle e^{2i(\theta-\hat\theta)}\rangle|^{-2}-1.
\end{equation}
The factor of two in the exponent takes care of this invariance at
the expense of having a phase in the range $[0,\pi)$.
Accordingly, the fidelity reads
\begin{equation}
    \label{eq:FidSVS}
    F=|\langle e^{2i(\theta-\hat\theta)}\rangle|,
\end{equation}
and  the relation between $V$ and the statistical variance
$(\Delta\hat\theta)^2$ is now
\begin{equation}
    \label{eq:VFD}
    V\simeq 2(1-F)\simeq 4(\Delta\hat\theta)^2.
\end{equation}

{}From Eq.~\eqref{eq:FidSVS}, and assuming that $\theta$ is uniformly distributed in the interval
$[0,\pi)$, one can easily obtain that the bound
to the fidelity is formally equivalent to Eq.~\eqref{F-coherent}:
$F\leq \sum_k|\rho_{k\,k+1}|$, where here $\rho$ is written in the
basis of symmetrized states $\ket{2k}$ defined in
Eq.~\eqref{eq:sym-squeezed} as $\rho_{k k'}=\bra{2k} \rho \ket{2k'}$.
Likewise a measurement that saturates this bound is given by
\begin{equation}
\label{eq:generalMeas-squ}
    [O_\phi]_{k l}=\frac{1}{\pi}e^{-2i\phi(k-l)}
\end{equation}
(also written in the same basis), and the optimal estimator is
$\hat\theta_\phi=\phi$, with $\phi\in[0,\pi)$. Taking into account
Eq.~\eqref{eq:statePhi}, the explicit form of the bound reads
\begin{eqnarray}
    \label{eq:GeneralFid}
    F\leq (1-\beta^2)^{N/2}\sum_{k=0}^\infty\beta^{2k+1} h_k h_{k+1}.
\end{eqnarray}
Using the techniques shown in Appendix~\ref{Fsvs}, this expression
can be recast into an integral form which is much easier to study,
both analytically and numerically. It reads
\begin{equation}
    F=\frac{N}{2}\beta (1-\beta^2)^{N/2}
        \int_{0}^1
   \frac{ du\, u^{\frac{N-2}{4}} I_0\left(\tfrac{2-N}{4}\log
   u\right) }{(1-u\beta^2)^{N/2+1}} \ ,
   \label{eq:SVSIntegral-text}
\end{equation}
where $I_0(x)$ is the zero order modified Bessel
function~\cite{tables}. Eq.~\eqref{eq:SVSIntegral-text}  follows
from Eq.~\eqref{eq:SVSIntegral} upon changing variables from $\tau$
to $u=(1-\tau)/\beta^2$.

Eq.~\eqref{eq:SVSIntegral-text} enables us to compute the Holevo
variance in a very efficient way for arbitrary values of $N$ and
$n$. Fig.~\ref{fig:loglog} shows a log-log plot of $NV$ against $n$ for $1\leq
N\leq 9$. We see that for low $n$ all curves have the same slope,
while for large $n$, as we anticipated in the introduction, the
slopes increase up to $N=4$ and then stabilize. We next proceed to
calculate these two scalings analytically.
\subsection{Large squeezing}
\label{sec:N8} 
\begin{figure}[t]
    \includegraphics[width=.49 \textwidth]{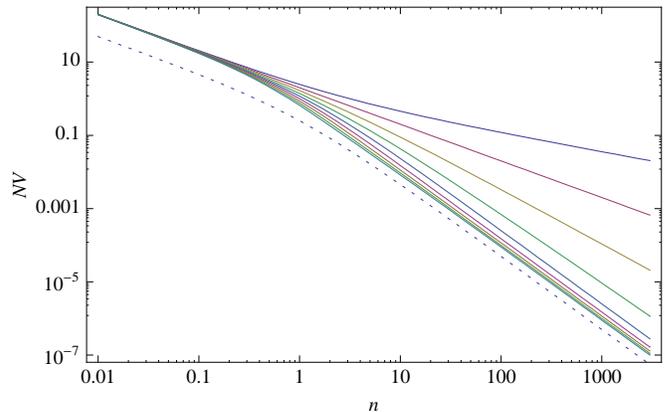}
    \caption{Log-log plot of the scaled phase variance $NV$ defined in Eq.~\eqref{eq:SVSvar} for $N$ copies of SVS, $1\leq N \leq 9$ (top to
    bottom),
    as a function of the mean photon number $n$ (solid lines).
    For large $n$ the lines become steeper as $N$ increases, in agreement with Eq.~\eqref{eq:varresults}. The slopes stabilize for $N\geq5$.
    For small $n$ the slopes are independent of $N$, as Eqs.~\eqref{eq:small-squeezing} and~\eqref{variance-N} show. The dotted line is the limiting curve ($N\rightarrow\infty$)  for both small and large $n$. It follows from the HL~\eqref{variance-N}, as discussed in Sec.~\ref{discussion}.}   \label{fig:loglog}
\end{figure}

\begin{figure}[t]
    \includegraphics[width=.49 \textwidth]{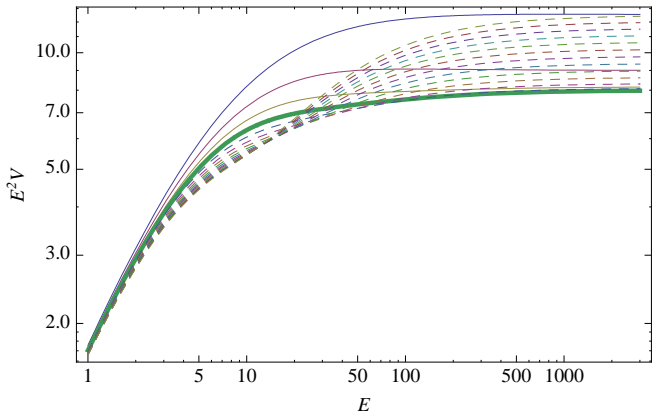}
    \caption{Log-log plot of the rescaled phase variance
    $E^2 V$ as a function of the total available energy
    $E=nN$ (in units of $\hbar \omega$).
    The thin solid lines correspond (from top to bottom)
    to $N=5,6,7$; the thick green line to $N=8$.
    Dashed lines correspond to $N\geq9$ in increasing
    order from bottom to top on the right side of the plot.
    For large $E$, variances scale as $E^{-2}$ (all lines have
    horizontal asymptotes) and $N=8$ clearly provides
    the smallest variance.}   \label{fig:N8}
\end{figure}

For very energetic SVS, $n\gg1$
$(\beta\rightarrow1)$, the phase can be estimated with arbitrary
accuracy; $\lim_{\beta\rightarrow 1}F=1$. In this regime the crucial
issue is to know the rate at which perfect estimation is achieved.
For the sake of readability the technical details of the calculation
are collected in Appendix~\ref{Fsvs}. In this section we only
summarize and comment the main results. {}From
Eqs.~\eqref{eq:FN-squeezed} and~\eqref{eq:VFD} one readily sees that
the statistical variance at leading order in $1/n$ is
\begin{eqnarray}
    \label{eq:varresults}
    (\Delta\hat\theta)^2=\left\{
    \begin{array}{ll}
        \displaystyle {\xi_N\over 2n^{N/2}} &(N\leq3)\\[1em]
        \displaystyle {(1/8)\log n+\xi_4\over 2n^2}\qquad &(N=4)\\[1em]
        \displaystyle {1\over8(N-4) n^2} & (N\geq5),
    \end{array}
    \right.
\end{eqnarray}
where the values of $\xi_N$ are given in Table~\ref{tab:coeff}. We
recall that these statistical variances are obtained assuming that one
can perform the most general (collective) measurement on the
$N$-copy state~$\rho_{\theta}^{\otimes N}$. We find it remarkable
that the $n^{-2}$ scaling is only achieved for $N\geq5$ (see
Fig.~\ref{fig:loglog}).
\begin{table}[b]
\begin{tabular}{c| c c c c }
\hline \hline
$N$ & 1 & 2 & 3 & 4 \\
\hline
$\xi_N$& $\ 0.55\ $ & $\ 1/2\ $ & $\ 0.58\ $ & $\ 0.30\ $ \\
\hline\hline
\end{tabular}
\caption{Values of the coefficients~$\xi_N$ in Eq.~(\ref{eq:varresults}).} \label{tab:coeff}
\end{table}
Notice that for a single copy, the optimal scaling is $n^{-1/2}$, as
compared to $n^{-2}$, which one would naively expect from HL. This
scaling  is  even worse than that attained with coherent states
$n^{-1}$ [see Eq.~\eqref{F-coherent}].   The same $n^{-1/2}$ scaling shows up in the $n$-photon two mode state discussed
in~\cite{berry}.
Using our analysis we can also study the optimal splitting of energy
among copies. If one has a fixed, but large, amount of energy $E$
that can be divided among $N$ copies, each with mean photon number
$n=E/N$, it is straightforward to obtain from Eq.~\eqref{eq:varresults}
that asymptotically the optimal choice is $N=8$. This result is also
clear from Fig.~\ref{fig:N8}.

\subsection{Small squeezing}
Let us now briefly focus on the low energy regime. This regime may
be relevant in practical situations where the amount of squeezing is
bounded by technological limitations. For $\beta\ll1$ we have
$\beta\simeq\sqrt n$ and we can easily find the leading behavior of~$F$ by
keeping the first terms in  Eq.~\eqref{eq:GeneralFid}. For $n N\ll 1$, a simple
calculation gives
\begin{equation}\label{eq:small-squeezing}
    F\simeq\sqrt{\frac{nN}{2}} 
    ; \quad
    V\simeq\frac{2}{nN}
    ,
\end{equation}
where the approximate expression for~$V$ follows from~(\ref{eq:SVSvar}) and~(\ref{eq:FidSVS}).
This shows that the scaling is independent of~$N$, as is clear from
Fig.~\ref{fig:loglog}. Note also that the fidelity only depends on
the total energy $E=nN$.
\section{Individual measurements: Large $N$}\label{sec:individual}
We have shown that the accuracy that can be achieved with a single
SVS increases with $n$ at a much lower rate than that inferred from
the HL. The natural question that arises is whether the latter can
be achieved by performing the same measurement on a large set of
identical copies. In other words,  we wish to know if the
attainability of the HL requires some sort of classical
communication between measurers dealing with the various copies. The
Fisher information allows to address this question. In
single-parameter estimation the Fisher information provides an
asymptotic bound (the Cram\'{e}r-Rao bound) for the accuracy that
can be attained by repeating the same measurement on each copy of
the sample.
 Recall that the  Fisher
information is defined as~\cite{CT}
\begin{equation}\label{eq:fisher}
  I(\theta)=  \int d\phi~p(\phi|\theta) \left(\frac{\partial \log p(\phi|\theta)}{\partial
  \theta}\right)^2,
\end{equation}
where $p(\phi|\theta)$ is the conditional probability of obtaining
the outcome $\phi$ upon measuring on the state $|\psi_\theta\rangle$
that carries a phase $\theta$ . In our case the measurement is
defined by Eq.~\eqref{eq:generalMeas-squ} and the corresponding
outcome probabilities are
\begin{equation}\label{eq:prob-fisher}
    p(\phi|\theta)=|S(\phi-\theta)|^2,
\end{equation}
where
\begin{equation}\label{eq:stheta}
   S(\phi)=\left(\frac{1-\beta^2}{\pi^2}\right)^{1/4}\sum_k \frac{\sqrt{(2k)!}}{k!}
    \left(\frac{\beta e^{2i\phi}}{2}\right)^k .
\end{equation}
The covariance of the measurement~\eqref{eq:generalMeas-squ} implies
that the Fisher information is independent of $\theta$, as is
apparent from the form of the conditional
probability~\eqref{eq:prob-fisher}. For highly energetic SVS
($\beta\simeq1$) we get
\begin{equation}
\label{eq:fisher-2} I\simeq{6\over7(1-\beta)^2}\simeq\frac{24}{7}n^2+\dots
\end{equation}
(we refer  to Appendix~\ref{sec:fisher} for details), and by using
the Cram\'er-Rao bound $N (\Delta\hat\theta)^2\geq
I^{-1}$, we obtain
\begin{equation}\label{eq:variance-asymptotic}
   (\Delta\hat\theta)^2\geq \frac{7}{24}\frac{1}{n^2N}.
\end{equation}
Recall that this bound is attainable for a large sample, $N\to
\infty$, with, e.g., a Bayesian or maximum likelihood
estimator~\cite{braunsteincaves94}.

Hence, we obtain that the optimal measurement performed sequentially
on a large sample of identically prepared SVS gives an accuracy that
scales as the HL, a result that could not easily be anticipated.
Notice, however,  that the coefficient of the variance is more than
a factor of two larger than that of the HL (7/24 as compared to
1/8). This means that in order to obtain the optimal accuracy we may
require  classical communication~\cite{amonras06}, i.e.,
adaptivity of the measurements.

\section{Discussion}\label{discussion}



{}From Eq.~\eqref{eq:varresults} and Fig.~\ref{fig:loglog} we see
that the HL, which predicts a scaling $n^{-2}$ for the variance, is
only achieved for $N\geq5$. Had we naively extrapolated the scaling
$n^{-N/2}$  for $N\leq 4$, we would have predicted a breakdown of
the HL (see Appendix~\ref{Fsvs}). Of course, this is not the case:
for $N>4$ the terms with the dangerous exponent behavior in
Eq.~\eqref{eq:SVSIntegral} become sub-dominant in the large $n$
limit, and the statistical variance scales as dictated by the HL.

We can also obtain the exact dependence in the mean number of photons for large $N$.
The easiest way is to rewrite $h_k h_{k+1}$ in Eq.~\eqref{eq:GeneralFid} as
$h_k^2\sqrt{(N/2+k)/(k+1)}$, so that
\begin{equation}\label{eq:fid-average}
    F=\beta\sum_{k=0}^{\infty}\wp_k \sqrt{\frac{N/2+k}{k+1}},
    \end{equation}
 where the negative binomial probability distribution $\wp_k$ is defined in Eq.~\eqref{eq:negative-binomial}. To compute this expectation value we
expand the square root  around the mean of the distribution, $\langle k\rangle =(N/2) \beta^2/(1-\beta^2)$, up to second order. By recalling that $(\Delta k)^2=\langle k\rangle/(1-\beta^2)$, it is straightforward to obtain
\begin{equation}
	F=1-\frac{(\beta^2-1)^2}{4\beta^2N},
\end{equation}
or equivalently
\begin{equation}
    \label{variance-N}
    (\Delta\hat\theta)^2=\frac{1}{8n(n+1)N},
\end{equation}
which is the exact expression of the HL (see~\cite{yurke86,
amonras06}\footnote{Note the wrong sign in Eq. (10.39)
of~\cite{yurke86}} and references therein).

Our results have a number of implications both of fundamental and of
practical interest. In the high energy regime they show that,
somewhat unexpectedly, the phase resolution of a single squeezed
vacuum state is worse than that of a coherent state with the same
energy. Our results indicate that whenever a phase measurement is to
be performed through interaction with a single copy (or mode) of the
probe, one is better off using coherent states. This scenario
changes significantly as one moves to the multi-copy case. For two
copies the resolution is already comparable to that of coherent
states and the variance reaches the HL scaling ($n^{-2}$) for $N>4$.
It is important to notice that we have considered arbitrary
collective measurements and, hence, our results give the ultimate
precision bounds allowed by quantum mechanics. In addition, our
analysis provides the means to determine the optimal energy
splitting among copies, showing that for highly energetic SVS, the
optimal choice is $N=8$, as shown in Fig.~\ref{fig:N8}.

In the low energy regime we have shown that the variance (fidelity) is only a function
of the total available energy, regardless the way it splits
among copies. This is relevant for practical implementations of
squeezed state metrology, where usually the amount of available
squeezing is small.

We have also analyzed the asymptotic accuracy of Holevo's canonical
phase measurement~\eqref{eq:generalMeas-squ} when it is performed on
each copy of the sample. It gives rise to a scaling of the variance
which agrees with the HL of SVS up to a constant multiplicative
factor of the order of two.
In some sense, this result relaxes the need for adaptive protocols
at the expense of having an asymptotic rate that is roughly a half
that of the optimal protocol. This shows that individual
non-adaptive measurements can harness the enhanced phase variance
offered by SVS.

\section{Acknowledgments}
We are grateful to A.~Ac\'in, J.~Calsamiglia, M.~Mitchell and  M.B.~Plenio for
very useful discussions, and to D.~Diego, for contributing to the
very early stages of this work. We acknowledge financial support
from the Spanish MCyT, CONSOLIDER2006-00019 and FIS2005-01369, and
CIRIT 2005SGR-00994. A.M. also acknowledges financial support from
QIP IRC GR/S82176/01.
%
%
\appendix
\section{Calculation of {\boldmath $F$} for SVS}
\label{Fsvs} In this Appendix we provide details of our calculation
of the fidelity for SVS. Our starting point is
Eq.~\eqref{eq:GeneralFid}. Taking into account Eq.~\eqref{eq:dks} we
can write
\begin{equation}\label{eq:fid-appendix}
       F\leq (1-\beta^2)^{N/2}\sum_{k=0}^\infty\frac{\beta^{2k+1} c_k}
       {\sqrt{(k+1)(k+N/2)}},
\end{equation}
where we have defined
\begin{equation}
c_k=\frac{N}{2} \binom{k+N/2}{k}.
\end{equation}
We next apply the integral representation \eqref{integral-trick-1}
to both $(k+1)^{-1/2}$ and $(k+N/2)^{-1/2}$,
\begin{multline}
\frac{1}{\sqrt{(k+1)(k+N/2)}}=\\ \int_0^\infty \int_0^\infty
 d t_1 d t_2\frac{\exp[-(k+1)t_1-(k+N/2)t_2]}{\pi\sqrt{t_1 t_2}}.
\end{multline}
This enables us to sum up the series (over $k$)
in~\eqref{eq:fid-appendix} by recalling the negative binomial
expansion $\sum_k c_k z^k =(N/2)(1-z)^{-N/2-1}$. We are left with a
double integral, which can be further simplified by the change of
variables~$(t_1,t_2)\rightarrow(u,\tau)$:
\begin{eqnarray}
    t_1&=&-u\log\left(\tfrac{1-\tau}{\beta^2}\right),\\
    t_2&=&-(1-u)\log\left(\tfrac{1-\tau}{\beta^2}\right),
\end{eqnarray}
and by recalling the integral representation of the zero order
modified Bessel function of the first kind~\cite{tables}
\begin{equation}
    I_0(x)=\frac{1}{\pi}\int_0^1 du \frac{e^{(1-2u)x}}{\sqrt{u(1-u)}}.
\end{equation}
The remaining integral can be cast as
\begin{eqnarray}
    \label{eq:SVSIntegral}
    F=\frac{N}{2}\beta(1-\beta^2)^{N/2}\int_{1-\beta^2}^1\frac{g(\tau,\beta)}{\tau^{1+N/2}}
    d\tau,
\end{eqnarray}
where
\begin{equation} \label{eq:gtau}
    g(\tau,\beta)=\frac{1}{1-\tau}
    \left(\frac{1-\tau}{\beta^2}\right)^{\frac{2+N}{4}}I_0\left(\tfrac{2-N}{4}
    \log\tfrac{1-\tau}{\beta^2}\right).
\end{equation}

Eqs.~\eqref{eq:SVSIntegral} and~\eqref{eq:gtau} provide a very
useful expression of $F$. The integral over $\tau$ can be computed
to arbitrary accuracy and it is valid for any number of copies and
for any average photon number. It allows, e.g.,  to find out the
optimal splitting of the available energy $E$ among copies for phase
estimation (see Sec.~\ref{sec:N8} and Fig.~\ref{fig:N8}).


We next derive from~\eqref{eq:SVSIntegral} the high energy scaling
of the fidelity. Let us first introduce some short-hand notation. By
$\mathcal{N}(\beta)$, we denote the prefactor
in~\eqref{eq:SVSIntegral},
\begin{equation}
    \mathcal N(\beta)=\frac{N}{2}\beta(1-\beta^2)^{N/2},
\end{equation}
and notice that it is of order $n^{-N/2}$. We further define
$S_P(\tau,\beta)=\sum_{k=0}^P G_k(\beta)\tau^k$ to be the truncated
Taylor expansion of $g(\tau,\beta)$ around $\tau=0$ up to order $P$.
One can easily check that its corresponding integral over $\tau$,
\begin{equation}
    \mathcal{G}_1(\beta)=\int_{1-\beta^2}^1\frac{S_{[N/2]}(\tau,\beta)}{\tau^{1+N/2}}d\tau ,
\end{equation}
is of order $n^{N/2}$ ($[x]$ is the integer part of $x$), whereas
the integral of the remainder,
\begin{equation}
 \mathcal{G}_2(\beta)=\int_{1-\beta^2}^1\frac{g(\tau,\beta)-S_{[N/2]}(\tau,\beta)}
 {\tau^{1+N/2}}d\tau ,
\end{equation}
is of order $n^0$. The leading contribution can be  computed by
taking the limit $\beta\rightarrow 1$ in $\mathcal{G}_2$. Thus, up
to subleading order, we have
\begin{equation}
    \label{eq:Fsplitting}
    F=\mathcal N(\beta)\left[\mathcal{G}_1(\beta)+\mathcal{G}_2(1)\right].
\end{equation}
We next expand the first term, $\mathcal
N(\beta)\mathcal{G}_1(\beta)$, in~\eqref{eq:Fsplitting} in powers of
$n^{-1/2}$. This expansion provides the leading order contribution
to the fidelity (the unity), a subleading term of order $n^{-2}$,
which becomes dominant only for $N>4$, and a series of inverse power
contributions starting at order $n^{-N/2}$. These latter become
irrelevant if $N>4$. On the other hand, the second term, $\mathcal
N(\beta)\mathcal{G}_2(1)$, results in corrections of order
$n^{-N/2}$ which contribute to the subleading order if $N\leq 4$ and
become irrelevant otherwise (of order smaller than $n^{-2}$).

In summary, we have
\begin{eqnarray}
    F=1-\left\{
    \begin{array}{ll}
        \displaystyle {\xi_N\over n^{N/2}} &(N\leq3)\\[1em]
        \displaystyle {(1/8)\log n+\xi_4\over n^2}\qquad &(N=4)\\[1em]
        \displaystyle {1\over4(N-4) n^2} & (N\geq5),
    \end{array}
    \right.
    \label{eq:FN-squeezed}
\end{eqnarray}
where  $\xi_N$ are formally defined in terms of definite integrals.
Its numerical values are given in Table~I. The exact value
$\xi_2=1/2$ can be trivially obtained from Eq.~\eqref{eq:fid-average},
which yields $F=\beta$.

\section{Phase estimation with individual measurements}
\label{sec:fisher}
In this Appendix we compute the Fisher information of the
optimal 1-copy measurement. This leads to the Cram\'{e}r-Rao bound
discussed in section~\ref{sec:individual}.

We first notice that the dominant behavior of $S(\theta)$ in
Eq.~\eqref{eq:stheta} is determined by the asymptotic expressions of
the factorials in the sum. The Stirling approximation gives
\begin{eqnarray}\label{eq:stheta-stirling}
 S(\phi)
   &\simeq &\left(\frac{1-\beta^2}{\pi^2}\right)^{1/4}
\frac{1}{\pi^{1/4}}\sum_k
   \frac{(\beta e^{2i\phi})^k}{k^{1/4}} \nonumber \\
&=&\left(\frac{1-\beta^2}{\pi^2}\right)^{1/4}
\frac{\mathrm{Li}_{1/4}(\beta e^{2 i\phi})}{\pi^{1/4}},
\end{eqnarray}
where $\mathrm{Li}_{s}(z)=\sum_{k=1}^{\infty} z^k/k^s$ is the
polylogarithm function of order $s$~\cite{tables}. Since we are only
interested in the asymptotic behavior for large squeezing ($\beta\to
1$), we can use the first order expansion around $z=1$, which reads
 $\mathrm{Li_{1/4}}(z)= \Gamma(3/4)
(1-z)^{-3/4}+\dots$~\cite{tables}. We note in passing that the
probability law obtained by retaining only this term is consistent
with the normalization condition,  i.e., $\int p(\phi|\theta) d\phi
\to 1$ when $\beta \to 1$. In this limit we can easily compute the
derivatives required in~\eqref{eq:fisher} and obtain
\begin{equation}\label{eq:fisher-approx}
I \simeq \frac{9 \Gamma(\frac{3}{4})^2\sqrt{1-\beta^2}}{\pi^{3/2}}
\int_{-\frac{\pi}{2}}^{\frac{\pi}{2}}\! \frac{d\phi\,
\sin^22\phi}{(1-2\beta \cos2\phi+\beta^2)^{\frac{11}{4}}}\, ,
\end{equation}
where we have used the rotational invariance of the integral. Notice
that the Fisher information is independent of the phase of $\pt$.
This is just a consequence of  the continuous and covariant
character of the measurement~\eqref{eq:generalMeas} and the isotropy
of the prior. Again, in  the limit~$\beta\to 1$ the main
contribution to the integral is peaked around $\phi\to 0$ and we can
Taylor-expand the trigonometric functions and safely extend the
integration limits from $(-\pi/2,\pi/2)$ to $(-\infty, \infty)$. We
thus obtain
\begin{eqnarray}\label{eq:fisher-approx2}
 && \kern-3em I\simeq\frac{36
\Gamma^2(\frac{3}{4})\beta^2\sqrt{1-\beta^2}}{\pi^{3/2}}
\int_{-\infty}^{\infty} \frac{d\phi\, \phi^2 }{[(1-\beta)^2
+4\beta \phi^2]^{\frac{11}{4}}}\nonumber \\
&& \kern-3em \phantom{I}= \frac{3}{7}\frac{\sqrt{2
\beta(1+\beta)}}{(1-\beta)^2}\simeq \frac{6}{7(1-\beta)^2}.
\end{eqnarray}
This is the Fisher information used in Eq.~\eqref{eq:fisher-2} in
the main text.



\end{document}